# Multidimensional Range Queries on Modern Hardware

Stefan Sprenger
Humboldt-Universität zu Berlin
Berlin, Germany
sprengsz@informatik.hu-berlin.de

Patrick Schäfer
Humboldt-Universität zu Berlin
Berlin, Germany
schaefpa@informatik.hu-berlin.de

Ulf Leser
Humboldt-Universität zu Berlin
Berlin, Germany
leser@informatik.hu-berlin.de

## ABSTRACT

Range queries over multidimensional data are an important part of database workloads in many applications. Their execution may be accelerated by using multidimensional index structures (MDIS), such as kd-trees or R-trees. As for most index structures, the usefulness of this approach depends on the selectivity of the queries, and common wisdom told that a simple scan beats MDIS for queries accessing more than 15%-20% of a dataset. However, this wisdom is largely based on evaluations that are almost two decades old, performed on data being held on disks, applying IO-optimized data structures, and using single-core systems. The question is whether this rule of thumb still holds when multidimensional range queries (MDRQ) are performed on modern architectures with large main memories holding all data, multi-core CPUs and data-parallel instruction sets.

In this paper, we study the question whether and how much modern hardware influences the performance ratio between index structures and scans for MDRQ. To this end, we conservatively adapted three popular MDIS, namely the R*-tree, the kd-tree, and the VA-file, to exploit features of modern servers and compared their performance to different flavors of parallel scans using multiple (synthetic and real-world) analytical workloads over multiple (synthetic and real-world) datasets of varying size, dimensionality, and skew. We find that all approaches benefit considerably from using main memory and parallelization, yet to varying degrees. Our evaluation indicates that, on current machines, scanning should be favored over parallel versions of classical MDIS even for very selective queries.

## 1 INTRODUCTION

Multidimensional range queries (MDRQ) are selection queries that specify a query interval for some or all dimensions of a multidimensional data space. MDRQ are an extremely common part of many workloads; important application areas include:

**OLAP.** Data warehouses use a multidimensional data model. Queries against this model very often result in MDRQ [17, 27]. For instance, a query may ask for sales in a certain price range within a certain period and within a certain range of products.

**Sensor Data.** Recent technologies like Internet of Things or cyber-physical systems are based on large numbers of sensors, which typically measure multiple features of their surveilled entities [25]. Often, only entities characterized by feature values within certain intervals are of interest for further analysis, like all locations within certain ranges regarding temperature, humidity, average and peak wind speed, and sunshine intensity.

**Genomics.** Precision medicine is largely based on analyzing the mutational landscape of entire populations or disease cohorts. Individual mutations are described using a broad range of features, like genomic location, functional impact, known disease associations, or properties of the person where it was found [16]. Researchers analyze these data using MDRQ, searching for instance all mutations in the coding regions of a certain cluster of genes present in patients of a certain age and weight range who suffer from a certain disease [24, 38]. Characteristic for these data and workloads are (a) a large number of tuples, (b) a moderate number of dimensions (3-100), and (c) range predicates over all or some of these dimensions[1].

MDRQ can be answered either by scanning all data or by employing multidimensional index structures (MDIS), of which quite a number have been proposed over the last decades [3, 6, 12, 15, 41, 42]. For historical reasons, most MDIS were designed for machines that feature single-core CPUs and small main memory capacities leading to single-threaded execution and IO reduction as major design goals. For such systems, Weber et al. [41] have shown that sequential scans should be favored over MDIS when roughly 20% or more of all indexed data need to be visited, assuming that accessing consecutive blocks (as in a scan) is at least 5 times faster than random access (as necessary for most MDIS). It seems that since then, this threshold has been used as the basic rule of thumb for choosing access paths in MDRQ. However, the capabilities of typical database servers have changed considerably over the last decades. Modern hardware features particularly interesting for MDRQ are large main memory capacities, multi-core and multi-threaded machines, and SIMD instructions: (1) Main memory has grown so large that most applications can hold all their data in memory, using disk only for persistence and fault tolerance [11, 19]. (2) Modern CPUs support many ways of parallelization, e. g., thread-level parallelism [23], or data-level parallelism [32]. Furthermore, we see a change in the typical workloads applied on MDIS. All use cases described above are analytical, i.e., their predominant access operations are reads. Data are updated rarely, essentially never deleted, and inserts are almost always performed in bulk [34]. In the database community, the shift to analytical workloads handled in main memory resulted in increased popularity of a column-wise storage layout [37]; in contrast, the classical MDIS were designed for row-wise data layouts. Thus, it is time to re-evaluate the performance of MDIS for MDRQ to see if the traditional rule of thumb still holds. Clearly, such a re-evaluation requires an adaptation of the original index structures to the features of modern hardware (see Figure 1) and should be carried out using analytical workloads.

---
[1] Queries over high-dimensional datasets or using similarity predicates are out of scope of this work; for supporting such use cases, we refer the reader to excellent surveys, like [6].

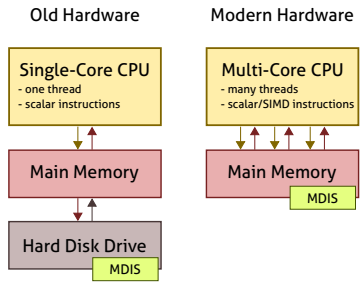

**Figure 1: Classical disk-based set-up for MDIS (left) versus an adaptation to modern hardware (right).**

In this experimental analysis, we study the question whether and how much the changes in hardware and workloads influence the performance of MDIS compared to sequential scans. To this end, we adapted three popular MDIS to be executed in a parallel and in-memory setting, namely (1) the R*-tree [2], an optimized variant of the R-tree [15], (2) the kd-tree [3], an index structure already originally designed for in-memory computations, and (3) the VA-file [41], which can be considered as a mixture between a MDIS and a sequential scan. Our adaptation is conservative in the sense that we withstood the temptation to design new, highly-tuned parallel MDIS, because our aim is to evaluate the classical approaches, which are still in use quite a lot and also included in many mature database systems, e. g., SQLite (R*-tree[2]), PostgreSQL (R-tree[3], kd-tree[4]), or Oracle (R-tree[5]). Our aim is to propose techniques, where existing implementations can be re-used with minimal adaptation, thus raising the practical relevance of our results.

We compare the adapted MDIS to the performance of a parallel, in-memory scan-based MDRQ implementation using both real-world and synthetic analytical workloads over datasets of varying size, dimensionality, and skew. Again, we employ only simple parallelization schemes and refrain from using highly-tuned scan implementations, like BitWeaving [26]. For providing a more realistic evaluation, we also devise a novel MDRQ benchmark derived from a genomic use case, consisting of eight parameterized classes of queries over a real-world dataset from the 1000 Genomes Project [8], consisting of 10 Million data points with 19 dimensions. This benchmark as well as all implementations, other workloads and datasets are freely available on our website[6].

Our results indicate that, although all approaches benefit considerably from modern hardware features, this effect is much more pronounced for simple scans than for MDIS. Accordingly, the rule of thumb to choose between a traditional MDIS and a scan based on query selectivities of MDRQ should be adjusted: from the classical 15%-20% down to around 1%.

## 2 BACKGROUND AND FOUNDATIONS
### 2.1 Problem Definition

We study the performance of multidimensional range queries (MDRQ). A dataset $D = \{t_0, t_1, ..., t_{n-1}\}$ is a collection of $n$ data objects sharing the same attributes. Each data object $t_i = \{v_0, v_1, ..., v_{m-1}\}$ is a sequence of $m$ attribute values. Alternatively, a data object can be described as a $m$-dimensional feature vector belonging to $\mathbb{R}^m$. In the following, we use the terms attribute, feature, and dimension synonymously.

We assume that an MDRQ returns the unique identifiers of all matching data objects. A *complete-match* multidimensional range query $MDRQ(D) = \{p_0, p_1, ..., p_{m-1}\}$ selects all qualifying data objects from $D$. For each dimension $j$, $MDRQ$ specifies a predicate $p_j = [lb_j, ub_j]$ that consists of a lower ($lb_j$) and upper ($ub_j$) boundary. A data object $t_i$ qualifies for $MDRQ$ iff all predicates are evaluated to true, i.e., $\forall j : lb_j \leq t_i[j] \leq ub_j$. MDRQ that specify predicates for only a subset of all dimensions are called *partial-match* queries. A partial-match query can be interpreted as a complete-match query that uses the predicate $[-\infty, +\infty]$ for all dimensions that are not queried.

The selectivity $sel(MDRQ)$ of a range query is defined as the percentage of data objects from $D$ that match the query. Queries that select only a small (large) portion of all data are considered to have a high (low) selectivity [36]. In addition to $sel(MDRQ)$, we define $sel_j(MDRQ)$ as the percentage of data objects from $D$ that match $MDRQ$ for dimension $j$. If all dimensions are statistically independent from each other, $sel(MDRQ) = \prod_{j=0}^{m} sel_j(MDRQ)$. For most real-world datasets this equation does not apply due to correlated dimensions [14].

### 2.2 Multidimensional Index Structures

According to [13], MDIS can be divided into point access methods (PAM) and spatial access methods (SAM). PAM are used to search sets of multidimensional data points, whereas SAM may also store objects with spatial extensions, such as rectangles. Nevertheless, SAM are frequently used to store points, as we do in this paper. To study whether and how much the features of modern hardware change the performance characteristics of MDIS, we choose three specific MDIS we consider as representative for their classes. These are the R*-tree (SAM), the kd-tree (PAM), and the VA-file. The R*-tree and the kd-tree are tree-based index structures, which recursively partition the space or dataset using certain split strategies. In contrast, the VA-file is a mixture between a flat partitioning index, like the grid file [30], and a sequential scan. Decades after their original publication, these three MDIS can still be considered as the state of the art for range-querying multidimensional data and are widely used in popular DBMS (see Introduction).

Defining these MDIS years after their invention is not as straightforward as one might think, as many variations have been proposed, sometimes with identical and sometimes with different scopes. In the following, we describe the specific variations we use in this paper which are all very close to the published methods; their adaptations to modern hardware will be explained in the subsequent sections. For completeness, at the end of the section, we also describe our baseline implementation of scan-based MDRQ.

---
[2] https://sqlite.org/rtree.html
[3] https://www.postgresql.org/docs/9.6/static/xindex.html
[4] https://www.postgresql.org/docs/9.6/static/spgist.html
[5] http://docs.oracle.com/html/A88805_01/sdo_intr.htm
[6] https://www2.informatik.hu-berlin.de/~sprengsz/mdrq

### 2.2.1 R-tree/R*-tree.
The R-tree [15] is a SAM that manages multidimensional data objects in a balanced tree of hierarchically organized minimum bounding rectangles (MBR). A MBR provides the minimal enclosure of a set of objects. In an R-tree, leaf nodes are blocks of data objects while inner nodes hold MBR enclosing all objects of their respective subtree. The R-tree was designed as a disk-based MDIS storing all data on disk. Thus, node sizes are adjusted to disk page sizes such that one node fits into one disk page. For executing a MDRQ, the search algorithm starts at the root node and hierarchically traverses the tree down to the leaf nodes. At each inner node, the algorithm intersects the query object with the node's MBR to determine those subtrees that may include data objects matching the query; other subtrees are pruned. Whenever a leaf node is reached, all data objects matching the query are added to the result set.

The performance of an R-tree largely depends on the number of subtrees that must be visited, which again depends on the way how objects are placed within the tree during their insertion. The placing of data objects is determined by a split strategy which decides how to proceed whenever a leaf node overflows after insertion. The split strategy of the R-tree led to many overlapping MBR and thus rather slow retrieval times. The R*-tree [2] is an optimized variant of the R-tree that aggressively re-inserts data objects when a node overflows leading to reduced MBR overlaps and faster queries at the cost of higher build-up cost. Since we focus on analytical workloads, we chose the R*-tree as basis for our adaptations.

### 2.2.2 kd-tree.
The concept of a kd-tree [3] is similar to that of a binary search tree, but it indexes $k$-dimensional objects. Every node of a kd-tree holds a data object. Inner nodes split the data space into two parts according to a delimiter dimension and a delimiter value. For a delimiter dimension $d_j$, the left subtree holds all data objects having a smaller or equal value than the inner node's data object in this dimension, and the right subtree holds all data objects having a greater value. Accordingly, every inner node could be understood as an axis-aligned hyperplane whose dimensionality depends on the depth of the node. When executing a range query, the search algorithm recursively traverses the tree from the root node to the leaf nodes. At each visited node, two actions are conducted. First, the stored data object is added to the result set if it is contained in the query range. Second, the delimiter dimension of the stored data object is compared with the corresponding dimension of the query object to determine which subtree needs to be taken to continue searching. Note that, unlike for point queries, MDRQ usually have to visit multiple parallel subtrees of a kd-tree.

We chose to use the original kd-tree (and not the kdb-tree [33] which is more suited for classical, IO-based RDBMS) as basis for our adaptation because it is an in-memory data structure by design. Like for R-trees, the shape of the kd-tree is determined by the split strategy used during insertions and, in turn, determines the performance of queries. The main decision to take upon an insert is the choice of the delimiter dimension of the next (new) inner node. As in the original paper, we choose the delimiter dimensions in a round-robin fashion, which promises a robust behavior over a wide range of data distributions.

### 2.2.3 VA-file.
The VA-file [41] is a blend of a flat space-partitioning index, like the grid file, and a sequential scan. It divides the $m$-dimensional data space into $2^b$ rectangular cells. Each cell is associated with a bucket that holds the actual objects. Data objects are inserted by hashing their coordinates to obtain the respective bucket; in the literature, this process is also called "approximation". The $b$ bits are distributed over the $m$ dimensions by assigning $b_j$ bits to dimension $d_j$ such that $b = \sum_{j=0}^{m} b_j$, and the hash functions per dimension should be chosen such that data objects are distributed evenly over the buckets. Buckets are stored on disk, whereas pointers to the buckets are managed in an in-memory array of size $2^b$. When executing a range query on a VA-file, the search algorithm approximates the query object, determines all buckets that have a non-empty intersection, and scans those buckets to find all objects contained in the query.

As in the original proposal, we implement the VA-file as a non-adaptive index. This implies that we statically assign the number of bits per dimension, using $b_j = 2$ for each dimension, and choose the hash function per dimension by partitioning the range of values evenly into $2^2 = 4$ intervals. Note that this index can still be updated, yet it would become unsuitable if future data objects follow data distributions grossly different than the ones initially used for determining the hash functions.

### 2.2.4 Sequential Scan.
We compare the performance of MDIS for MDRQ to that of a scan over the entire dataset. Listing 1 shows the non-parallel version of a sequential scan which we use as basis for all adaptations. Using a *for* loop, the search algorithm iterates over all $n$ data objects stored in an array named *data*. It compares each data object with the range query object defined by the parameters *lower* and *upper*. If all $m$ dimensions match, the object identifier (implicitly defined by the array index $i$) is added to the result set.

This simple implementation leaves room for further optimizations. First, the algorithm could compare the dimensions in the order of the expected selectivities, i.e., querying highly selective dimensions first, leading to an earlier break of the inner loop. This would, however, require estimates about the selectivity of single-dimension range queries. Second, we use the same algorithm for partial-match and for complete-match queries, although for partial-match queries, the comparisons with the ranges of the unspecified dimensions always return TRUE. We could skip these comparisons. However, we chose to use this unoptimized implementation because our intention is to compare a *standard* scan with different *standard* MDIS. If we tuned our scan, we would also have to tune our MDIS, which in turn would make existing implementations unusable.

## 3 PARTITIONING FOR PARALLELIZATION
As we will describe in detail in Section 5, we parallelize the scan and MDIS implementations by partitioning the data. We use and evaluate two different partitioning schemes, namely horizontal partitioning and vertical partitioning.

Figure 2 illustrates the techniques when used to divide 20 5-dimensional data objects into 5 partitions.

### Listing 1: Multidimensional range scan.

```
std::vector<int>
mdrq(int[][] data, int n, int m, int[] lower, int[] upper) {
  std::vector<int> results;
  for (int i = 0; i < n; ++i) {
    bool match = true;
    for (int j = 0; j < m; ++j) {
      if (data[i][j] < lower[j] || data[i][j] > upper[j]) {
        match = false; break;
      }
    }
    if (match)
      results.push_back(i);
  }
  return results;
}
```

## 3.1 Horizontal Partitioning

Horizontal partitioning divides a dataset of $n$ objects into $p$ partitions. Each partition holds a near-identical number of data objects, i.e., partition size $\approx n/p$. To obtain a robust load balancing, we assign data objects at random to partitions and set $p = t$ given that $t$ is the number of available hardware threads (virtual cores). Such a scheme is simple to implement and maintain, also in the presence of inserts or updates[7]. All approaches for MDRQ we consider in this work treat the different partitions as independent (which, for MDIS, means that actually $p$ indexes are built), which allows using existing implementations without any adaptations. At search time, each partition is assigned to one CPU thread that produces a partial result; once all threads have finished, these results are simply concatenated to produce the final result set. Horizontal partitioning is thus similar to the approach proposed as Parallel VA-file [40], except that here we target multi-threading provided by modern CPUs instead of distributing queries to multiple physical machines.

A major advantage of horizontal partitioning is its simplicity. Partitions are chosen at random which means that no further data structures are necessary for managing them. Creation of the overall result set requires only concatenating partial result sets, incurring only minimal synchronization effort. Also the number of partitions could be chosen freely to diminish the effect of stragglers, if present. At the downside, for partial-match queries, whole data objects have to be considered and must always be read entirely into the CPU cache, although only a subset of all dimensions are queried.

## 3.2 Vertical Partitioning

Vertical partitioning slices data objects along their dimensions, creating one partition for each dimension. This scheme is similar to the tree striping technique presented in [5] and also followed in column stores, which have become quite popular for analytical workloads in database systems [37]. In this scheme, the number of partitions $p$ is fixed and always equal to the dimensionality $m$ of the data. For query execution, an $m$−dimensional range query is split into $m$ one-dimensional range queries, which are executed in parallel by distinct threads. Accordingly, also the degree of parallelism in this phase is fixed at $m$. Eventually, the one-dimensional results must be intersected to compute the final result set. In our implementation, we use a bitmask of size $n$ for each partition to note for each data object whether it matches in this dimension or not. Once the bitmasks have been computed, we intersect them using a bitwise AND operator to determine the final bitmask reflecting all dimensions. This intersection is performed in parallel on different partitions of the bitmasks.

[7]Assuming that the workload follows certain known distributions, one could also think of assigning data objects to partitions following a non-random fashion to arrive at an optimal load balancing, as, for instance, described in [4]. Such a scheme would have to be implemented individually for each approach considered. Again, we refrain from applying such optimizations to keep our comparison fair.

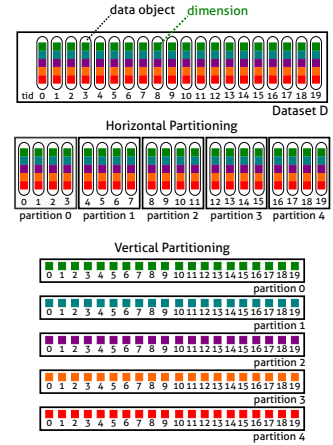

Figure 2: Horizontal and Vertical Partitioning used to divide 20 5-dimensional data objects into 5 partitions.

Clearly, vertical partitioning is only meaningful for our scan-based MDRQ implementation, as MDIS would degenerate to one-dimensional indexes for the partitions that are created. The main advantage of vertical partitioning is its built-in support for partial-match queries, as only those partitions have to be accessed that are referred to in the query. For complete-match queries, it offers no advantage over horizontal partitioning in terms of data access, as all values have to be accessed in both cases. It has the general disadvantage that it requires large intermediate data structures (the bitmasks) and also a more complex way of producing the final result set. It is also rather complicated to achieve a good load balancing when $m < t$, since threads remain idle, or $m > t$, because then not all partitions can be processed concurrently and some partitions are scanned after other partitions. In both cases, one should start to chop dimensions into subsets and parallelize on this finer level of granularity. Again, we do not tune our implementation in such manners but stick to the simple scheme of assigning one thread to each partition.

## 4 VECTORIZED INSTRUCTIONS

Modern CPUs support vectorized instructions, which process multiple values with a single instruction by using specialized registers, a feature usually called Single Instruction Multiple Data (SIMD). As opposed to multi-threading, which enables thread-level parallelism, vectorized instructions enable data-level parallelism, where the degree of parallelism depends on the width of the specialized registers[8]. When working on a data type for which $k$ values fit into

[8]We use the AVX instruction set through Intel's Intrinsics.

**Listing 2: Vectorized comparison of an m-dimensional MDRQ search object with an m-dimensional data object.**

```
1   bool match = true;
2   __m256 lower_reg, upper_reg, search_reg, lower_res, upper_res;
3   int i, mask_lower, mask_upper, mask;
4   const int compares = (m / 8) * 8;
5   for (i = 0; i < compares; i += 8) {
6     lower_reg  = _mm256_loadu_ps(&lower[i]);
7     upper_reg  = _mm256_loadu_ps(&upper[i]);
8     data_reg   = _mm256_loadu_ps(&data[i]);
9     lower_res  = _mm256_cmp_ps(lower_reg, data_reg, _CMP_LE_OQ);
10    upper_res  = _mm256_cmp_ps(upper_reg, data_reg, _CMP_GE_OQ);
11    mask_lower = _mm256_movemask_ps(lower_res);
12    mask_upper = _mm256_movemask_ps(upper_res);
13    mask       = mask_lower & mask_upper;
14    if (mask < 0xFF) {
15      match = false; i = m; break;
16    }
17  }
18  
19  for (; i < m; ++i) {
20    if (data[i] < lower[i] || data[i] > upper[i]) {
21      match = false; break;
22    }
23  }
24  
25  if (match)
26    // add data object to result set
```

these registers, SIMD offers a theoretical speed-up of $k$; however, this value is rarely achieved in practice as multiple other factors, such as memory bandwidth and the concrete instruction to perform, play an important role [32]. For instance, AVX instructions, which work on 256-bit SIMD registers, can process eight 32-bit floating-point values in parallel with one instruction and offer a theoretical speed-up of a factor of 8. The details of how we use SIMD instructions are discussed in the next section. However, we apply SIMD instructions to the basic operation of comparing a MDRQ search object (with up to $m$ dimensions) to a data object (with $m$ dimensions), as shown in Listing 2 for the case that all dimensions are of type *float*. Obviously, vectorizing this operation is only beneficial if $m \geq k$.

In the beginning (see Lines 1 to 4), SIMD registers and helper variables are initialized. On the machine we use for evaluation, SIMD registers are 256 bits wide, which means that eight floating-point values can be processed in parallel. We therefore process all values in chunks of eight (see Lines 5 to 17) and the remaining values in a separate loop (see Lines 19 to 23). The variable *compares* holds the number of to-be-executed SIMD comparisons. After loading chunks of eight dimensions of the data object, the query's lower boundary and the query's upper boundary into SIMD registers, the algorithm compares the lower boundary (*lower_reg*) and the upper boundary (*upper_reg*) with the data object (*data_reg*) using SIMD instructions. Results of the comparisons are stored in the variables *mask_lower* and *mask_upper*. SIMD-based comparisons return bitmasks that indicate if a comparison, e. g., less or equal (*_CMP_LE_OQ*), was successful. If all comparisons were successful all bits are set to 1, which equals to $0xFF$ in our case (see Line 14). Otherwise, we can abort (see Line 15) and prune further comparisons. Finally, if all dimensions of the data object match the given search query, we add it to a result set (see Lines 25 to 26).

## 5 CONSERVATIVE ADAPTATION OF MDIS TO MODERN HARDWARE

In this section, we describe our conservative adaptations of the R*-tree, the kd-tree, and the VA-file to compare their performance in executing MDRQ with that of a parallel scan using horizontal partitioning or vertical partitioning. In particular, we describe our changes to the original data structures performed to adapt (1) to main-memory storage, (2) to the availability of multiple threads, and (3) to SIMD instructions. As our evaluation focuses on analytical workloads, we only discuss the search algorithms. In all cases we tried to keep the original code untouched as much as possible to allow re-usage of existing, proven implementations to the largest possible degree.

Our overall strategy to adapting MDIS is to (a) keep the original storage layout but hold all blocks in main memory, (b) partition the data horizontally at random into almost-equally sized chunks and build one MDIS per instance which work in parallel during MDRQ, and (c) use SIMD only for the most time-consuming operation, i.e., matching of data objects with the query. This strategy has the advantage that it could also be applied to any other MDIS, as we essentially build a conventional MDIS for every partition; only the orchestration of the different MDIS (which does not require any synchronization), the object matching using SIMD, and the result concatenation for obtaining the final result set have to be added. Furthermore, load balancing is quite simple as it only requires adapting the size and number of partitions, which also suffices to adapt to the number of available hardware threads on a machine.

On the downside, this scheme performs many comparisons redundantly, as every partition-wise MDIS has to cover the entire space, making pruning of entire MDIS instances impossible. Accordingly, for the hierarchical MDIS (R*-tree, kd-tree) every search initially traverses similar layers, while for the VA-file multiple buckets with the same approximation have to be searched. Note, however, that this redundant work is performed in parallel. Eliminating these redundant comparisons would require multiple threads to operate on the same index structure, leading to wide-spread changes specific for each MDIS. Especially for the hierarchical MDIS, such a strategy would also have to solve complex load balancing problems, as the number of branches grows exponentially with the depth of the tree, creating over-provisioning in the upper levels and under-provisioning in the lower levels. Such schemes are beyond the scope of our work but have been explored, for instance, in [22] or [35].

### 5.1 R*-tree

We base our implementation of the R*-tree (see Section 2.2.1) on the open-source library libspatialindex[9]. We performed the following adaptations: (1) We keep all nodes of the tree in main memory. We do not adjust the node sizes to the sizes of disk blocks any more. Instead, we slightly adjust the default values of the used library from a capacity of 100 to 96 data objects (MBR) for leaf nodes (inner nodes), such that nodes are perfectly aligned with cache lines regardless of the dimensionality of the dataset[10]. (2) We horizontally partition the data at random into almost-equally sized chunks and build one

---

[9]https://libspatialindex.github.io/
[10]The used implementation stores dimension data as 8-byte double values. Hence, choosing a capacity that is a multiple of 8 always aligns nodes to 64-byte cache lines.

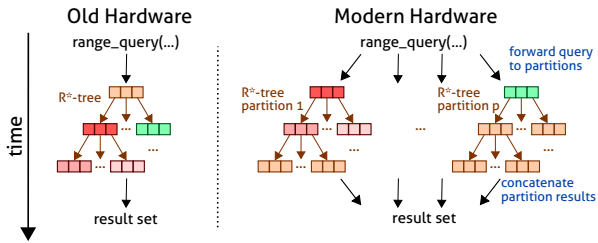

Figure 3: Sequential execution of a MDRQ on a single-threaded R*-tree vs. parallel execution of a MDRQ on p instances of an R*-tree, where each instance manages the data of a certain partition (horizontal partitioning) and is searched with one thread (p threads in total).

R*-tree instance per partition. These are searched in parallel and the (partial) result sets are concatenated once all R*-tree instances have finished their search. We chose the number of partitions to be equal to the number of hardware threads on the evaluation machine. (3) We use Listing 2 to compare the query object with MBR in inner nodes. Since the used R*-tree implementation works with 64-bit floating-point values, we can compare only four values (instead of eight) with one SIMD instruction. Figure 3 illustrates the differences between sequentially querying a conventional R*-tree and executing a parallel MDRQ following our approach.

### 5.2 kd-tree

We implemented a kd-tree from scratch following the original proposal (see Section 2.2.2) with the following adaptations to exploit modern hardware features: (1) The data layout is already designed for main memory, thus no adaptations were necessary in this regard. (2) For using multiple threads, we use exactly the same method as for the R*-tree. (3) We exploit SIMD instructions as shown in Listing 2 for intersecting the query object with the data object in all nodes of the kd-tree.

### 5.3 VA-file

We also implemented the VA-file from scratch using the following adaptations to the original method as described in Section 2.2.3: (1) All buckets are stored in main memory. The bucket size is derived from the length of the approximations; based on preliminary experiments, we use 2 bits per dimension for the approximations with values roughly evenly dividing the space per dimension, leading to partitions holding approx. $n/(4^m)$ data objects. (2) Again, we horizontally partition all data at random and build one VA-file per partition. (3) We use the algorithm from Listing 2 to compare the query object to all data objects from buckets whose approximations intersect the approximated query.

### 5.4 Parallel Scan on Horizontally Partitioned Data

For scanning horizontally partitioned data, we (1) hold partitions in $p$ independent $m$-dimensional arrays, where we set $p$ to the number of available hardware threads. (2) These arrays are concurrently scanned using a multidimensional range scan (see Listing 1). (3) During this scan, every data object is matched to the query object using SIMD instructions. Therefore, we replace the inner loop of the multidimensional range scan (see Lines 7-16 from Listing 1) with the vectorized algorithm shown in Listing 2.

### 5.5 Parallel Scan on Vertically Partitioned Data

For scanning vertically partitioned data, we (1) hold all partitions in $m$ one-dimensional arrays, each storing the value of one dimension per data object. (2) These arrays are concurrently scanned during an MDRQ; for partial-match queries, only the dimensions addressed in a query are accessed. Note that in this approach the degree of parallelism of a MDRQ is constrained by the number of dimensions specified in the query. Each partition creates one bitmask of size $n$, which are subsequently merged using efficient bitwise AND. Merging is performed in $t$ chunks of equal size. (3) In vertically partitioned data, only single dimensions of data objects are compared to the query object, hence the SIMD code from Listing 2 cannot be used. Instead, we vectorize each one-dimensional range scan similar to the approach proposed by Zhou and Ross [43].

## 6 GENOMIC MULTIDIMENSIONAL RANGE QUERY BENCHMARK (GMRQB)

The strength of any evaluation of MDRQ critically depends on the data and queries used. MDIS may perform very different for data following different distributions (uniform, unimodal or multimodal clustered, clustered in subspaces, etc.) and workloads of different characteristics (hot spot regions, partial- or complete-match queries, selectivities, etc.). While many evaluations of MDIS are performed only on synthetic data or synthetic workloads (such as [20], [31], or [39]), with the obvious advantages of being able to influence many parameters of the data and the workloads, we strived to also evaluate all our compared methods on real-world multidimensional data and workloads. To this end, we created a novel benchmark derived from genomics data for evaluating approaches to MDRQ. This section presents the GMRQ Benchmark, which consists of eight parameterized real-world (partial- and complete-match) MDRQ over a dataset of 10 Million genomic variations derived from the 1000 Genomes Project [8]. Thus, the benchmark consists of both real-world queries and real-world data. Data points are of moderate dimensionality (19 dimensions), and attributes feature very different numbers of distinct values. The dataset our benchmark builds upon is publicly available[11], which allows reproduction of our evaluation results and facilitates further research on MDRQ.

### 6.1 Variations and MDRQ

A human genome consists of approx. three Billion base pairs (the DNA) structured in 23 chromosomes. When sequencing a human, i.e., experimentally determining its genome, these three Billion base pairs are typically compared to a so-called human reference genome modeling a hypothetical "normal" human genome [18]. Deviations from this reference are typically called variations [7] (or mutations if they affect the human in some negative sense). On

---

[11] http://www.internationalgenome.org/data

Multidimensional Range Queries on Modern Hardware          preprint, 2018

| GMRQB Query Template | Average Selectivity | Average # of Queried Dimensions |
| --- | --- | --- |
| Query Template 1 | 10.76% ($\sigma$ = 7.24%) | 2 ($\sigma$ = 0.0) |
| Query Template 2 | 2.19% ($\sigma$ = 2.27%) | 5 ($\sigma$ = 0.0) |
| Query Template 3 | 5.36% ($\sigma$ = 3.61%) | 3 ($\sigma$ = 0.0) |
| Query Template 4 | 0.22% ($\sigma$ = 0.15%) | 4 ($\sigma$ = 0.0) |
| Query Template 5 | 0.20% ($\sigma$ = 0.15%) | 5 ($\sigma$ = 0.0) |
| Query Template 6 | 0.11% ($\sigma$ = 0.11%) | 6 ($\sigma$ = 0.0) |
| Query Template 7 | 0.05% ($\sigma$ = 0.06%) | 7 ($\sigma$ = 0.0) |
| Query Template 8 | 0.00001% ($\sigma$ = 0.00002%) | 19 ($\sigma$ = 0.0) |
| Mixed Workload | 1.58% ($\sigma$ = 3.58%) | 5.81 ($\sigma$ = 4.11) |

**Table 1: GMRQB query templates.**

average, a human genome features approx. 4-5 Million such variations [8]. Variations are not distributed at random over the genome, but certain regions are more prone to variations than others. For instance, coding regions, i. e., the forming genes, carry relatively few variations due to evolutionary pressure, whereas inter-gene regions are comparably variation-rich [21]. The premise of precision medicine[12] is to correlate an individual's variation profile to his or her susceptibility to diseases and treatment [28]. An important part of research in precision medicine is concerned with collecting large numbers of genomes together with medical information about the individual, and to perform statistical analysis of variation profiles regarding commonalities and differences between individuals w.r.t. health-related issues. In such studies, researchers routinely search for sets of variations sharing certain characteristics, e. g., being in the same genomic region, being present in the same class of diseases or a similar group of patients, being present in patients reacting in the same way to medication, etc. These searches eventually result in MDRQ on large variation databases, like [16].

## 6.2 The GMRQ Benchmark

The 1000 Genomes Project published the DNA of 2,504 human individuals from across the world to facilitate research in precision medicine and related areas [8]. The entire dataset includes 84.4 Million distinct variations grouped by individual and by genomic location. Each variation is characterized by a broad set of attributes, of which we use 19 for our benchmark dataset: *Chromosome* and *location* define the genomic position of the variation. *Quality*, *depth*, and *reference_genome* are meta data derived from the sequencing procedures. *Variation_id*, *allele_freq*, *allele_count*, *ref_base*, *alt_base*, *ancestral_allele*, and *variant_type* are meta data of the individual variation. Finally, attributes *sample_id*, *gender*, *family_id*, *population*, *relationship*, and *genotype* characterize the human being in which a particular variation was found. More details on this dataset can be found on our website.

The workload of the Genomic Multidimensional Range Query Benchmark (GMRQB) consists of eight realistic query templates, designed by a group of Bioinformaticians, that retrieve specific subsets of genomic variations interesting for further analysis. These queries specify range predicates on some, most, or all dimensions. Predicates may either specify single points or ranges of different sizes; for instance, we always use a point query for the attribute *gender*, yet always use ranges for the attribute *location*. All queries of GMRQB include predicates on the genomic location, i. e., attributes *chromosome* and *location*. Queries in the workload are templates

[12]See, for instance, https://allofus.nih.gov/.

where specific ranges have to be instantiated with meaningful values. For the genomic location, we use the RefSeq[13] database to align genomic ranges to coding regions and fill other variables using values found in the 1000 Genomes Project dataset. Our website provides all query templates of GMRQB. Table 1 shows the average selectivity and average number of queried dimensions of each query template. Except Query Template 8, all query templates are partial-match queries. As a concrete example, we list an instance of Query Template 3, which selects all variations from a certain *location* range on a certain *chromosome* which match the specified ranges on *quality*, *depth*, and *allele_freq*:

```
SELECT * FROM variations
  WHERE chromosome = 5
    AND location BETWEEN 100000 AND 1000000
    AND quality BETWEEN 10 AND 100
    AND depth BETWEEN 10 AND 1000
    AND allele_freq BETWEEN 0.5 AND 1;
```

## 7 EVALUATION

The objective of our evaluation is to investigate the performance of MDRQ on modern hardware. To this end, we compare three MDIS and two scan variants, which we adapted as described in Section 5, using synthetic and real-world workloads on synthetic and real-world datasets.

### 7.1 Experimental Setup

*7.1.1 Hardware.* We executed all experiments on a server equipped with two Intel Xeon E5-2620 CPUs (2 GHz clock rate, 6 cores, 12 hardware threads) and 32 GB RAM. In total, the machine features 12 cores and 24 hardware threads (hyperthreads). The CPU supports AVX instructions on 256-bit SIMD registers.

We also ran the experiments on another hardware platform to prove that our findings do not depend on the used hardware architecture. It features one Intel i7-5930K CPU (3.5 GHz clock rate, 6 cores, 12 hardware threads, AVX instructions) and 32 GB RAM. The experimental results are very similar on both platforms (data provided on our website).

*7.1.2 Methodology.* We present only experiments, where caches are warmed up (hot cache). All datasets are inserted in random order[14]. All experiments measure throughput, which is the number of operations, in our case MDRQ, each contestant can execute per second. Typically, we run a query workload consisting of 1,000 queries and measure the time $t_s$ each contestant needs to execute all queries. Then we divide the number of executed queries by $t_s$ to get the throughput.

*7.1.3 Competitors.* We include the R*-tree [2], the kd-tree [3], the VA-file [41] and a parallel scan over horizontally and vertically partitioned data in our experiments. All contestants are evaluated with and without multi-threading and SIMD instructions. All MDIS employ horizontal partitioning, which allows to use the same algorithms as for a single-threaded search. Unless otherwise noted, we set $p = t$ for horizontal partitioning to exploit all available processing units and because we do not expect any stragglers that would benefit from using $p > t$.

[13]https://www.ncbi.nlm.nih.gov/refseq/
[14]Note that hierarchical MDIS show an improved performance on non-uniform distributions given that data are inserted in sorted order and queries follow the same distribution (results not shown).

| Dataset | Data Objects | Dimensions | Domain per Dimension (real numbers) | Distinct Values per Dimension | Raw Dataset Size (MB) |
|---|---|---|---|---|---|
| **SYNT-UNI** (uniform distribution) | 10k | 5 | [0,1] | 9,950 (avg) | 0.19 MB |
| | 100k | 5 | [0,1] | 95,175 (avg) | 1.91 MB |
| | 1M | 5-100 | [0,1] | 632,257 (avg) | 19.07 MB - 381.47 MB |
| | 10M | 5 | [0,1] | 999,956 (avg) | 190.74 MB |
| **SYNT-CLUST** (with clusters) | 1M | 5 | [0,1] | 632,047 (avg) | 19.07 MB |
| **POWER** | 10k | 3 | [2556001,2566000]; [12857,17281]; [14142,19278] | 10,000; 627; 698 | 0.11 MB |
| | 100k | 3 | [2556001,2656002]; [12466,18247]; [13698,20395] | 100,000; 2,089; 2,290 | 1.14 MB |
| | 1M | 3 | [2556001,3556003]; [12466,18770]; [13698,20704] | 1,000,000; 4,325; 4670 | 11.44 MB |
| | 10M | 3 | [2,9875683]; [12282,24623]; [13281,26879] | 9,875,681; 6,840; 7,634 | 114.44 MB |
| **GMRQB** | 10M | 19 | Our website provides a detailed description of all dimension data of GMRQB. | | 724.79 MB |

Table 2: A description of the datasets used in our experiments.

*7.1.4 Software.* All software was implemented in C++11 and was compiled with GCC 4.8 using optimization flag *-O3*. We use an open-source thread pool library[15] to enable the reuse of POSIX threads. The R*-tree is based on the open-source implementation libspatialindex. On our website, we describe how to apply SIMD instructions to the used R*-tree implementation. For the remaining contestants, we use our own implementations.

## 7.2 Experimental Data and Workloads

We evaluate MDRQ on four different datasets. Table 2 provides the number of data objects, number of dimensions, domains of each dimension, distinct values of each dimension (for synthetic data, we provide average values over all dimensions), and raw dataset size. For all datasets, we use 32-bit floating-point values to manage dimension data. The used R*-tree implementation uses 64-bit floating-point values. When comparing the memory consumption of the contestants, MDIS need between 2.5 and 5.4 times more space than a sequential scan (data provided on our website). For the datasets SYNT-UNI, SYNT-CLUST and POWER, we use synthetic workloads containing only complete-match MDRQ. For GMRQB, we execute both complete-match and partial-match MDRQ.

*7.2.1 SYNT-UNI.* Synthetic data facilitates experiments with arbitrary dataset sizes (10k to 10M data objects) and dimensionalities (5 to 100 dimensions). For SYNT-UNI, we generate uniformly distributed data objects within the domain [0, 1]. For most experiments with SYNT-UNI, we generate MDRQ by randomly choosing two objects from the generated data and use those as lower/upper boundary. This results in queries with varying selectivities.

*7.2.2 SYNT-CLUST.* In contrast to the uniformly distributed SYNT-UNI, the 5-dimensional dataset SYNT-CLUST features between 1 and 20 clusters. It is well known that MDIS struggle with data, where the uniform assumption does not hold [10]. For SYNT-CLUST, we used a data generator provided by Müller et al. [29]. Within each cluster, data are uniformly distributed. We generate query workloads using the same technique as for SYNT-UNI.

*7.2.3 POWER.* The real-world dataset POWER is obtained from the DEBS 2012 challenge[16]. As in previous studies [39] with this dataset, we index three dimensions. As for the synthetic datasets, we generate MDRQ by randomly choosing two data objects from POWER and use those as lower/upper boundary.

*7.2.4 GMRQB.* We index genomic variation data provided by the 1000 Genomes Project. Using our own data importer, we transform raw variation data into 19-dimensional feature vectors (data objects). Attributes originally stored as strings, like the population of a sample, are transformed into floating-point values by hashing. Our website provides a detailed description of all attributes including the domain and number of distinct values[17]. For GMRQB, we use eight realistic MDRQ templates. We also evaluate a mixed workload that consists of all query templates randomly mixed together.

## 7.3 Impact of Multi-Threading and Vectorization

Figure 4 shows the throughput of MDRQ with an average selectivity of 0.1% ($\sigma$ = 0.002%) on 1M uniformly distributed data objects of moderate dimensionality (dataset SYNT-UNI) depending on the used hardware features. For all contestants, we evaluate a single-threaded (baseline) implementation, a single-threaded implementation exploiting SIMD instructions, a multi-threaded implementation, and a multi-threaded implementation using SIMD instructions.

*7.3.1 Multi-Threading.* First, we compare single-threaded to multi-threaded implementations (both without SIMD), where we used 24 software threads, which equals to the number of available hardware threads. Due to a dimensionality of $m = 20$, the scan with vertical partitioning uses only 20 threads. No contestant achieves a speedup factor near 24X. Apparently, the speedup is bounded by the number of physical cores, which equals to 12 in our evaluation machine. Hyper-threading is only beneficial for multi-threaded applications, where threads are frequently waiting for data to be loaded from main memory into CPU caches, making memory accesses the bottleneck; this is also described as "blocking". In contrast, for the highly-selective query workload considered here, most of the time the contestants are compute-bound. Even

---

[15]https://github.com/vit-vit/CTPL

[16]http://debs.org/?p=38

[17]https://www2.informatik.hu-berlin.de/~sprengsz/mdrq/#gmrqb

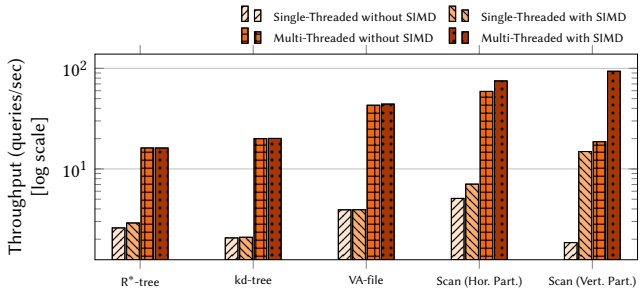

Figure 4: Throughput when executing MDRQ with an average selectivity of 0.1% on 1M 20-dimensional data objects depending on the used hardware features.

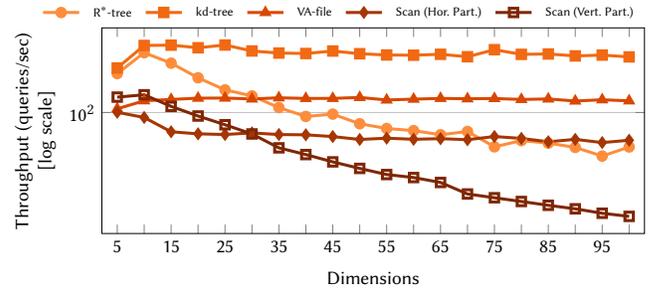

Figure 5: Throughput when executing range queries with an average selectivity of 0.4% (5 dimensions) to 0.0002% (> 10 dimensions) on 1M uniformly distributed data objects using 24 software threads depending on dimensionality.

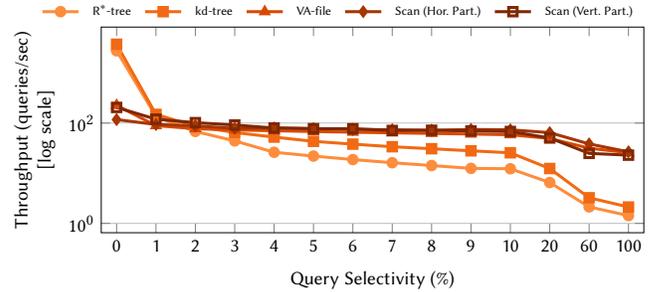

Figure 6: Throughput when executing range queries on 1M 5-dimensional uniformly distributed data objects using 24 software threads depending on query selectivity.

both scans and the VA-file are not memory-bound, because too many early breaks[18] occur.

*7.3.2 Vectorization.* The scan with vertical partitioning is the only contestant that benefits notably from vectorization, whereas MDIS and scans employing horizontal partitioning show almost no impact. When using SIMD instructions to compare MDRQ search objects with horizontally partitioned data objects, they have to compare at least eight dimensions before being able to prune further comparisons. In contrast, the scalar search allows early breaks as soon as the first mismatch occurs. However, for query workloads with a low selectivity and therefore fewer early breaks, the benefits from using SIMD instructions to process horizontally partitioned data increase (data not shown).

*7.3.3 Multi-Threading & Vectorization.* Although the scan with vertical partitioning shows comparatively small performance gains from multi-threading, when combining multi-threading with SIMD instructions it shows the largest speedup over a single-threaded scalar implementation among all contestants. In contrast, the remaining approaches, which all employ horizontal partitioning, benefit from multi-threading but show negligible performance gains when using SIMD instructions on top. For the subsequent experiments, we apply multi-threading and vectorization to all contestants.

### 7.4 Synthetic Data (Uniform Distribution)

*7.4.1 Dimensionality.* We measure the throughput of the contestants when executing MDRQ on 1M randomly generated data objects with different dimensionalities (5 to 100 dimensions). Clearly, the query selectivity increases with a growing dimensionality. The average query selectivity is 0.4% ($\sigma = 1.1\%$) for 5 dimensions, 0.002% ($\sigma = 0.01\%$) for 10 dimensions, and 0.0002% ($\sigma = 0.00003\%$) for more than 10 dimensions. Figure 5 shows the results.

The kd-tree achieves the highest throughput among all contestants regardless of the dimensionality. For up to 30 dimensions, the R*-tree shows the second best performance. For data with more dimensions, it is outperformed by the VA-file, which is less affected by the dimensionality of data. While the parallel scan employing horizontal partitioning shows a rather stable throughput independent from dimensionality, the performance of the scan using vertical partitioning decreases when dimensionality increases. In vertical partitioning, the number of partitions depends on the dimensionality of the feature space ($p = m$). The performance decreases because a growing number of partial result sets need to be managed and intersected when synchronizing CPU threads.

*7.4.2 Query Selectivity.* We measure the throughput of all contestants when executing MDRQ on 1M randomly generated 5-dimensional data objects following an uniform distribution w.r.t. query selectivity. Figure 6 shows the results.

For queries with a very high selectivity ($\leq$ 1%), the kd-tree shows the highest throughput and is closely followed by the R*-tree. For queries with a lower selectivity (> 1%), both parallel scans as well as the VA-file show the best performance and clearly outperform the hierarchical MDIS kd-tree and R*-tree. Overall, the performance of both scan variants and the (non-hierarchical) VA-file is very similar.

The throughput of all contestants decreases for queries with a lower selectivity for multiple reasons: (1) larger (partial) result sets need to be managed and synchronized impacting especially the scan on vertically partitioned data, (2) approaches employing horizontal partitioning can prune less dimensions when comparing MDRQ with data objects resulting in fewer early breaks, and (3) hierarchical MDIS (R*-tree and kd-tree) cannot prune subtrees but must visit the vast majority of tree nodes, which leads to lots of cache misses due to random access. The latter has the largest impact on the performance of MDRQ. Therefore, also on modern hardware, hierarchical MDIS remain very sensitive to query selectivity.

---
[18]When comparing a search object with a data object dimension by dimension, as soon as the first mismatch occurs, further comparisons can be pruned.

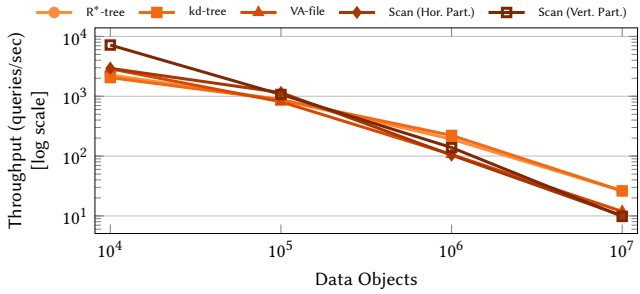

Figure 7: Throughput when executing range queries with an average selectivity of 0.4% on 5-dimensional uniformly distributed data objects using 24 software threads depending on the dataset size.

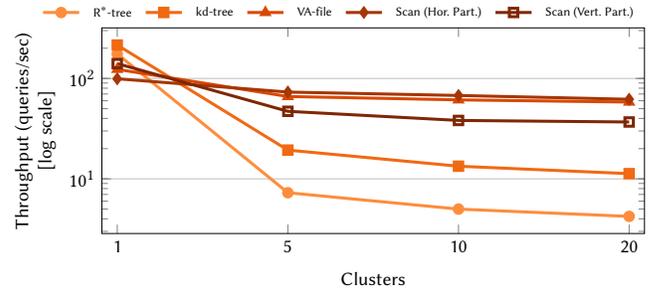

Figure 8: Throughput when executing range queries with an average selectivity of 0.38% (1 cluster) to 27.40% (20 clusters) on 1M 5-dimensional data objects using 24 software threads depending on the number of clusters.

*7.4.3 Dataset Size.* We measure the throughput when executing MDRQ with an average selectivity of 0.4% ($\sigma = 1.1\%$) on 5-dimensional data objects following an uniform distribution w.r.t. the number of objects. Figure 7 shows the results.

As expected, when the number of data objects increases, the search throughput of all contestants decreases because a growing number of data objects match the query. Interestingly, both parallel scans, especially the variant employing vertical partitioning, outperform MDIS for small datasets consisting of up to $10^5$ objects, although the query selectivity is very high. MDIS do not seem to be worthwhile for such small amounts of data. As the dataset size increases, the pruning techniques of MDIS pay off. MDIS can efficiently reduce the data space while the parallel scans have to consider all data objects for query evaluation.

## 7.5 Synthetic Data (Clusters)

We measure the throughput when executing MDRQ on 1M 5-dimensional data objects from the dataset SYNT-CLUST depending on the number of clusters. Recall that we are generating MDRQ by randomly picking two data objects as range boundaries. Thus, one MDRQ may cross several clusters, which results in a decreasing query selectivity as the number of clusters increases: 1 cluster (avg 0.38%, $\sigma = 0.94\%$), 5 clusters (avg 16.24%, $\sigma = 19.13\%$), 10 clusters (avg 23.12%, $\sigma = 21.88\%$), and 20 clusters (avg 27.40%, $\sigma = 22.71\%$). Figure 8 shows the results of this experiment.

Although the R*-tree and the kd-tree achieve the best throughput for the dataset with one cluster, their performance decreases as the number of clusters increases (which implies decreasing selectivities). In contrast, the VA-file and both parallel scans are less affected. Their throughput is almost independent of the number of clusters.

## 7.6 POWER

We measure the throughput of the contestants when executing MDRQ with an average selectivity of 11.12% ($\sigma = 13.43\%$) on the POWER dataset. In contrast to Figure 7, which shows the throughput for uniformly distributed data, Figure 9 visualizes the throughput on (skewed) real-world data of varying dataset size. This experiment confirms that the throughput of all contestants decreases when the number of data objects increases. As opposed to the experiments on synthetic data, scan-based approaches always outperform hierarchical MDIS regardless of dataset size.

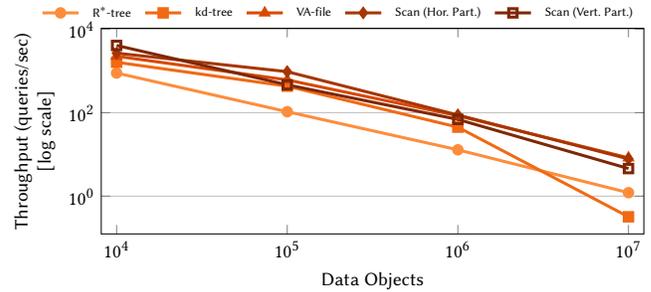

Figure 9: Throughput when executing range queries with an average selectivity of 11.12% on the 3-dimensional POWER dataset using 24 software threads depending on dataset size.

## 7.7 GMRQ Benchmark

We first study the throughput of all contestants when executing the query templates from the GMRQ Benchmark. Each template is instantiated 100 times using values derived from the 1000 Genomes Project dataset, as described in Section 6.2. Figure 10 shows the results of this experiment for each template and a mixed workload. Both parallel scans outperform all evaluated MDIS for the Mixed Workload, Query Template 1, Query Template 2 and Query Template 3. For Query Templates 4-8, which select only few data objects and have a selectivity (much) below 1%, especially the kd-tree shows its strengths and outperforms scanning.

In the next experiment, using the Mixed Workload, we evaluate the scalability of all contestants depending on the number of used threads. Note that the Mixed Workload of GMRQB consists of partial- and complete-match MDRQ that query 5.81 dimensions on average ($\sigma = 4.11$), which limits the potential benefits from multi-threading for vertical partitioning. Figure 11 shows the results.

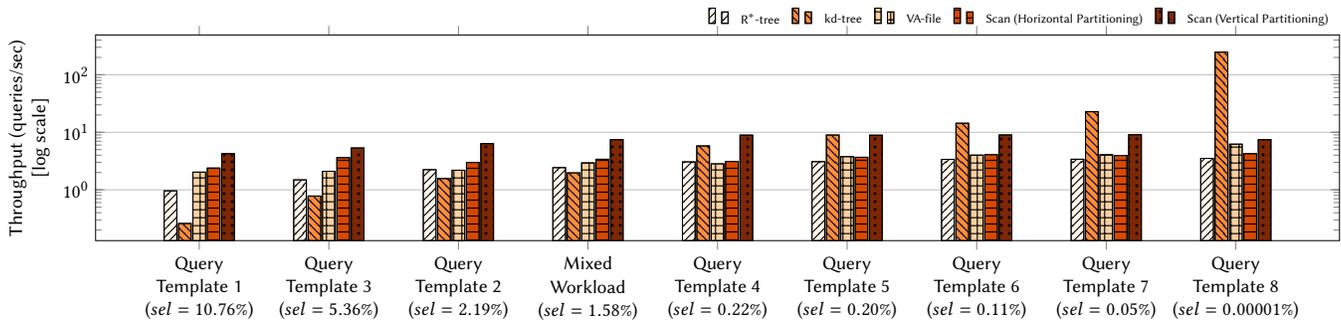

**Figure 10: Throughput of contestants when executing the GMRQB with varying selectivities on 10M 19-dimensional data objects from the 1000 Genomes Project dataset using 24 software threads (query templates are ordered by selectivity).**

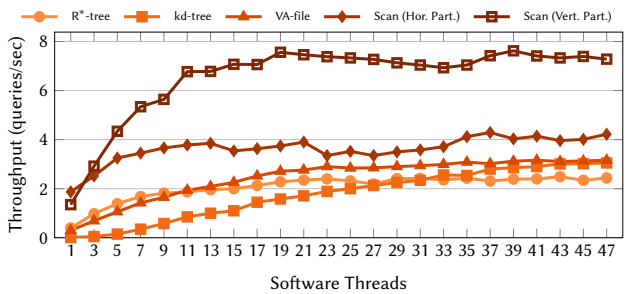

**Figure 11: Throughput of contestants when executing the Mixed Workload from GMRQB on 10M 19-dimensional data objects depending on the number of used software threads.**

For most contestants, the speedup from multi-threading is bounded by the number of physical cores. Confirming a memory access bottleneck, only the R*-tree and the kd-tree benefit from using more software threads than available physical cores, because hierarchical MDIS require many random accesses when evaluating queries having a moderate/low selectivity, which is in contrast to the highly-selective query workload from Section 7.3.1. When comparing the number of last level cache (LLC) misses[19] of single-threaded implementations it becomes clear that they are more frequently blocking than the scan-based approaches and therefore benefit from hyper-threading. For this workload, hierarchical MDIS show a 8X higher number of LLC misses than the VA-file and a 19X higher number of LLC misses than the sequential scan. Using more software threads than available virtual CPU cores (> 24) does neither yield performance benefits nor disadvantages.

## 8 SUMMARY AND DISCUSSION

Our comparison of two hierarchical tree-based MDIS (R*-tree, kd-tree), the VA-file, and two scan variants with different partitioning techniques after adapting them to the usage of main-memory storage, multi-threading and SIMD instructions allows for a number of interesting observations. As expected, MDIS in general excel for queries with very high selectivities as in such settings they can prune substantial parts of the search space and have to compare only a few data objects to the query. The main goal of our study was to re-evaluate the break-even point at which these advantages supersede the major disadvantage of MDIS, namely random access to the memory. Our experiments show that this point is surprisingly low, at around 1% selectivity, and thus much lower than the conventional rule-of-thumb, which targeted IO-based index structures. Although similar findings have been reported for one-dimensional range scans [9], to the best of our knowledge, we are the first to confirm these performance characteristics for the multidimensional case. Following the results of our study, scanning should be favored over indexing except for very selective queries. One should also note that highly selective queries are anyway very fast, regardless of the method used, which means that the absolute savings in time MDIS offer in such settings are very small. On top, scan-based MDRQ are much easier to handle, require no additional storage, are almost unaffected from the dimensionality of the data, lead to simple and effective load balancing, and offer predictable runtimes, which is a major plus when it comes to orchestrating the multiple operations of a complex analytical query.

---

[19] A last level cache miss occurs whenever data need to be transferred from main memory to the CPU caches.

There are also a number of further observations. Among the two hierarchical MDIS, in our evaluation the kd-tree clearly outperforms the R*-tree, which is not too much of a surprise as the R*-tree was originally designed to manage spatial objects and not points as was the case here. The VA-file offers almost never an advantage when compared to scans, but requires additional memory and synchronization. It only appears to be a sensible choice for data with very high dimensionality, in which case, however, also the memory requirements are large. When comparing the two scan methods, it seems that for complete-match queries the horizontal partitioning is preferable, whereas partial-match queries, especially if only a few dimensions are addressed, are handled more efficiently by scans over vertically partitioned data.

However, when considering our results one should always keep in mind that our adaptations to the indexes and scans we used were rather conservative. We already listed several ideas how the different methods could be further improved to take full advantage of modern hardware, like selectivity-based re-ordering of dimensions for matching data objects with the query, two-level partitioning for vertical scans for queries targeting less dimensions than available threads, adaptive hashing schemes for the VA-file, etc. Another obvious idea would be the usage of compression, especially for vertical partitioning [1]. Furthermore, we only looked at analytical workloads, although our implemented methods are all fully updateable; actually, we performed many experiments with different randomized insertion orders but could not observe any significant differences in runtimes (data not shown).

## ACKNOWLEDGMENTS

We would like to thank the bioinformaticians from our working group, especially Yvonne Lichtblau, for their valuable feedback on the design of the GMRQ Benchmark.

Stefan Sprenger is funded by the Deutsche Forschungsgemeinschaft through graduate school SOAMED (GRK 1651).